\documentclass[a4paper]{llncs}

\usepackage{url}
\usepackage{graphicx}
\usepackage{makeidx}  
\usepackage{wrapfig}
\usepackage[colorlinks=true,urlcolor=blue]{hyperref}

\begin{document}

\mainmatter  

\title{Development of Semantic Web-based Imaging Database for Biological Morphome}
\author{Satoshi Kume\inst{1,2,3} \and 
Hiroshi Masuya\inst{4,5} \and 
Mitsuyo Maeda\inst{2} \and
Mitsuo Suga\inst{2} \and\\
Yosky Kataoka\inst{1,2,3} \and 
Norio Kobayashi\inst{5,4,2}}  

\institute{RIKEN Center for Life Science Technologies (CLST), RIKEN,\\
6-7-3  Minatojima-minamimachi,  Chuo-ku,  Kobe,  Hyogo,  650-0047  Japan\\
satoshi.kume@riken.jp, kataokay@riken.jp
\and
RIKEN CLST-JEOL Collaboration Center, RIKEN\\
6-7-3  Minatojima-minamimachi,  Chuo-ku,  Kobe,  Hyogo,  650-0047 Japan\\
mitsuyo.maeda@riken.jp, msuga@jeol.co.jp
\and
RIKEN Compass to Healthy Life Research Complex Program, RIKEN Cluster for Science and Technology Hub, RIKEN,\\
6-7-1 Minatojima-Minamimachi, Chuo-ku, Kobe,  Hyogo,  650-0047 Japan\\
\and
RIKEN BioResource Center (BRC),\\
3-1-1, Koyadai,Tsukuba, Ibaraki, 305-0074 Japan\\
hmasuya@brc.riken.jp
\and
Advanced Center for Computing and Communication (ACCC), RIKEN,\\
2-1 Hirosawa, Wako, Saitama, 351-0198 Japan\\
norio.kobayashi@riken.jp
}
\maketitle 

\begin{abstract}
We introduce the RIKEN Microstructural Imaging Metadatabase, a semantic web-based imaging database in which image metadata are described using the Resource Description Framework (RDF) and detailed biological properties observed in the images can be represented as Linked Open Data. The metadata are used to develop a large-scale imaging viewer that provides a straightforward graphical user interface to visualise a large microstructural tiling image at the gigabyte level. We applied the database to accumulate comprehensive microstructural imaging data produced by automated scanning electron microscopy. As a result, we have successfully managed vast numbers of images and their metadata, including the interpretation of morphological phenotypes occurring in sub-cellular components and biosamples captured in the images. We also discuss advanced utilisation of morphological imaging data that can be promoted by this database.

\keywords{Microscopy ontology, Open Microscopy Environment, Imaging Metadatabase, Morphome, Morphological phenotype, Scanning electron microscopy}
\end{abstract}

\section{Introduction} 
Imaging data are an important fundamental aspect of contemporary life sciences. They facilitate the understanding of detailed morphological changes related to cellular functions and various diseases. We focus on microstructural images obtained using electron microscopes (EM), which provide detailed morphological information about tissues and cells at a nano-scale level \cite{Kas2015}. Typically, EM images have two significant problems. 1) Because the target area of an EM is quite small (sub-cellular level: less than a micrometre square), a wide-range overview (tissue level: millimetre square) is required to screen phenotypes. 2) Interpreting an EM image requires an advanced understanding of histology and histopathology.

To solve the first problem, we have developed a technology to obtain comprehensive microstructural imaging data of biotissues using scanning electron microscopy (SEM) with an autofocus function (Maeda, M. et al., under preparation). This imaging technology provides wide-range and high-resolution images of a sub-millimetre square area in biosamples. 
This type of comprehensive imaging is referred to as the morphome analysis, especially micro-morphomics in order to identify the totality of the microstructural morphological features and is considered to be an omic research. 
While such imaging data are accumulating rapidly as big data, a common procedure for sharing, integrating and analysing such data has not yet been developed. In addition, specialised experience is required to interpret morphological phenotypes in the images. Recently, Williams et al. launched the Image Data Resource (IDR), a prototype public database and repository for imaging data \cite{Williams2017}. The IDR is the first general repository for image data in life sciences. Thus, the potential impact of accessing, sharing and referencing imaging data in experimental results and published research has increased. However, issues associated with metadata, image formats and vocabulary associated with imaging experiments remain unresolved. To ensure reproducibility of experimental data and results, including expert image analysis, a common data publication framework and a multi-faceted ontology that describes various imaging experiments and experimental conditions are desired. Moreover, to complement specialised histology and histopathology experience and realise further automatic imaging data analysis, such as machine learning, metadata that describe related experimental conditions and phenotypes should be machine readable.

Our goal is to generate standardised machine-readable metadata that include related information, such as experimental conditions and phenotype data, for life sciences research. Currently, we are constructing an imaging metadata system for optical and electron microscopy that employs the Resource Description Framework (RDF). We are also developing an ontology to describe RDF metadata. Relative to the development of an ontology to describe imaging metadata, we have previously proposed general microscopy ontology concepts that involved translating XML-based Open Microscopy Environment (OME) metadata and the extension of incomplete vocabularies \cite{Kume2016}. The RDF ontology provides vocabularies and semantics to describe metadata, including information about electron microscopy imaging conditions, biosamples and sample preparation. In this paper, we describe the proposed RDF-based RIKEN Microstructural Imaging Metadatabase, which provides imaging metadata integrated with biosample and phenotype metadata and an imaging viewer that takes advantage of such metadata.
%
%
%
%
\section{Methods}
\subsection{Extension of Microscopy Ontology for Morphomics Data}
The OME is an open-source interoperable toolset for biological imaging data to manage multi-dimensional and heterogeneous optical microscopy imaging data \cite{Goldberg2005}. 
Previously, using the latest version of the OME data model, we translated the XML based OME data model into a base concept for an RDF schema \cite{Kume2016}. At this point, we identified missing concepts in the OME data model and extended the vocabularies to describe imaging metadata integrated with biosample and phenotype metadata.

To improve the ability to describe experimental situations, we further expanded vocabularies in the microscopy ontology. Note that phenotypic annotations often differ between observation of microscopic images and direct observation of biosamples because their magniﬁcation levels are different (i.e. sub-cellular and individual levels). Therefore, in the microscopy ontology, we enabled separate descriptions of phenotype data obtained from imaging analysis (i.e. sub-cellular phenotypes) and direct observation of biosamples (i.e. phenotype data of biosamples obtained from an individual).

In comprehensive microstructural imaging at the sub-tissue level, it is necessary to manage image data continuously photographed in units of several thousands of images or huge image data at several gigabyte levels using a tiling process. Recently, such huge image data have been converted and visualised using a pyramidal image format called the Deep Zoom Image (DZI) format. Therefore, we added new vocabularies (image directory, image overlap, tiling method, tiling map, etc.) to describe the metadata for tiling image data and DZI visualisation. In addition, the class of the DZI dataset is related to \texttt{rdfs:subClassOf} of the dataset class, and it has a structure that can link individual image data.

\subsection{Microstructural Bioimaging and Image Processing}
To prepare a biosample, we initially prepared wide-range ultra-thin (70 $\mathrm{nm}$) tissue sections obtained from a rat liver (600 $\mathrm{\mu m}$ x 500 $\mathrm{\mu m}$). Following the electron staining of the liver sections, we performed large-scale microstructural imaging at high spatial resolution (7.87 $\mathrm{nm/pixel}$) covering the entire area of the section using SEM. As a result, 288 16-bit images (5120 x 3840 pixels) were obtained, and the amount of data was approximately 10 GB.

To construct a tiling image, overlaps of resized images were computed using the Grid/Collection stitching plug-in in ImageJ/Fiji \cite{Preibisch2009} and stitched to a single tiling image. Then, the tiling image was converted to the DZI format using the Python-based code deepzoom.py (initially developed by Kapil Thangavelu) for visualisation by the electron microscopy viewer.

\subsection{Development of Electron Microscopy Viewer and the Annotation of Phenotype Data}
To display many microstructural images as a single large image, we developed the JavaScript-based RIKEN CLST Electron Microscopy Viewer, in which an interactive multi-resolution visualisation function was implemented using OpenSeadragon (\url{https://openseadragon.github.io}). 
In addition, overview, metadata linkage, snapshot, annotation, measurement and importing and exporting annotation file functions were implemented. The metadata of the viewer are a part of metadata based on the microscopy ontology and are contained in the taxon of National Center for Biotechnology Information (NCBI), animal strain, derives from, imaging method and staining method used to describe tiled images. We then annotated the phenotype data of the images, such as the binuclear cell phenotype (CMPO:0000213) \cite{Jupp2016} in the liver, using the electron microscopy viewer.

\subsection{Construction of Metadatabase for Microstructural Imaging Data of Biotissue}
RIKEN has been making efforts to publish metadata which is suitable for the cutting-edge research communities in life sciences by discussions with experts in various fields such as ontology, informatics and biology \cite{Kobayashi2016}. Furthermore, RIKEN has developed a database platform called the RIKEN MetaDatabase (\url{http://metadb.riken.jp}) \cite{Kobayashi2016} with the help of two virtual machines on the RIKENs private cloud platform using OpenLink Virtuoso [\url{https://virtuoso.openlinksw.com/}], and have already started conducting pilot operation as a metadata publishing service for biologists. In this study, we introduced an RDF-based metadatabase for large-scale tissue microstructural images. Then we generated the metadata and published the images and their metadata in the Imaging Metadatabase (\url{http://metadb.riken.jp/metadb/db/clstMultimodalMicrostruct}).

\section{Results and Discussion}
\subsection{Vocabulary Extension for Experimental Description and Morphome Data Description}

\begin{figure}[ht]
\centering
\vspace*{-2mm}
\includegraphics[width=10cm]{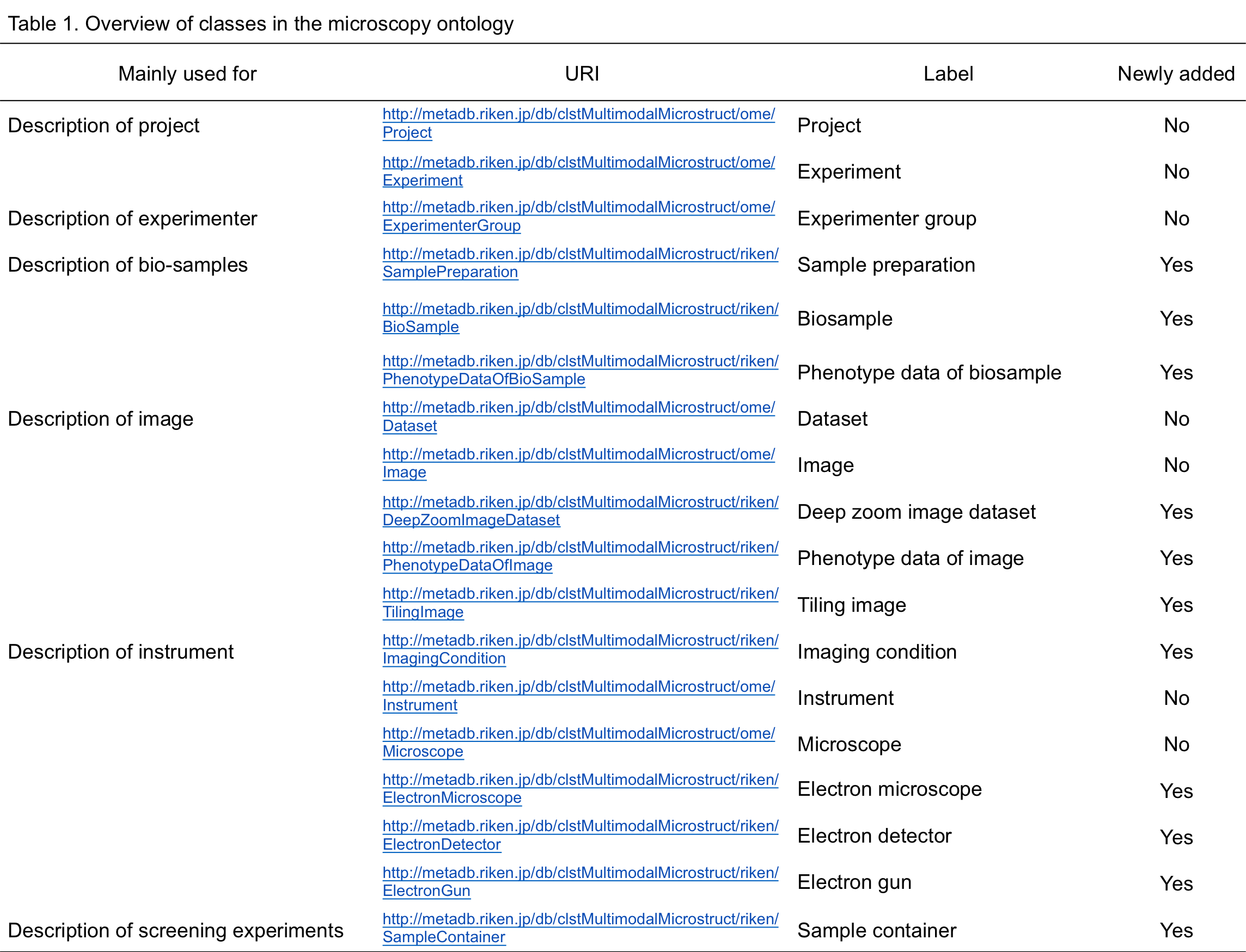}  
\label{fig:tab}
\end{figure}

We developed an ontology to define the upper-level concepts in microscope imaging \cite{Kume2016} as descriptions of ‘Project’, ‘Experimenter’, ‘Biosample’, ‘Image’, ‘Instrument’ and ‘Screening experiments’. This ontology is described in OWL (OWL 2 DL) and has 324 classes and 225 properties. The representative classes are shown in Table 1. For example, in the image descriptions, the classes, such as dataset and image, were translated directly from the OME data model. In addition, we added new classes, including a DZI dataset, phenotype data of biosample and image, bioformat and tiling image (Table 1). 

We defined a high-level concept so as to realise the common description of the description vocabulary of the microscopy imaging and the difference among various types of microscopies at the same time, by examining the concept in the microscopy ontology. As a specific example, in the microscopy imaging conditions, one of the difference between the optical microscope and the electron microscope is the apparatus and the initial parameters. Although they are individualised as subclasses, those that can be integrated using the higher concept ‘ImagingCondition’.

Additionally, we focused on the description of the phenotype data because phenotypic observations relating to the whole cell, cellular components and cellular processes are a noteworthy research topic in life sciences \cite{Jupp2016}. Then, we performed discrimination of the phenotype data relative to what was imaged (subject) and what was captured (image) as Phenotype data of sample and Phenotype data of image classes under the upper Phenotype data concept. 

\begin{figure}[ht]
\centering
\vspace*{-2mm}
\includegraphics[width=11cm]{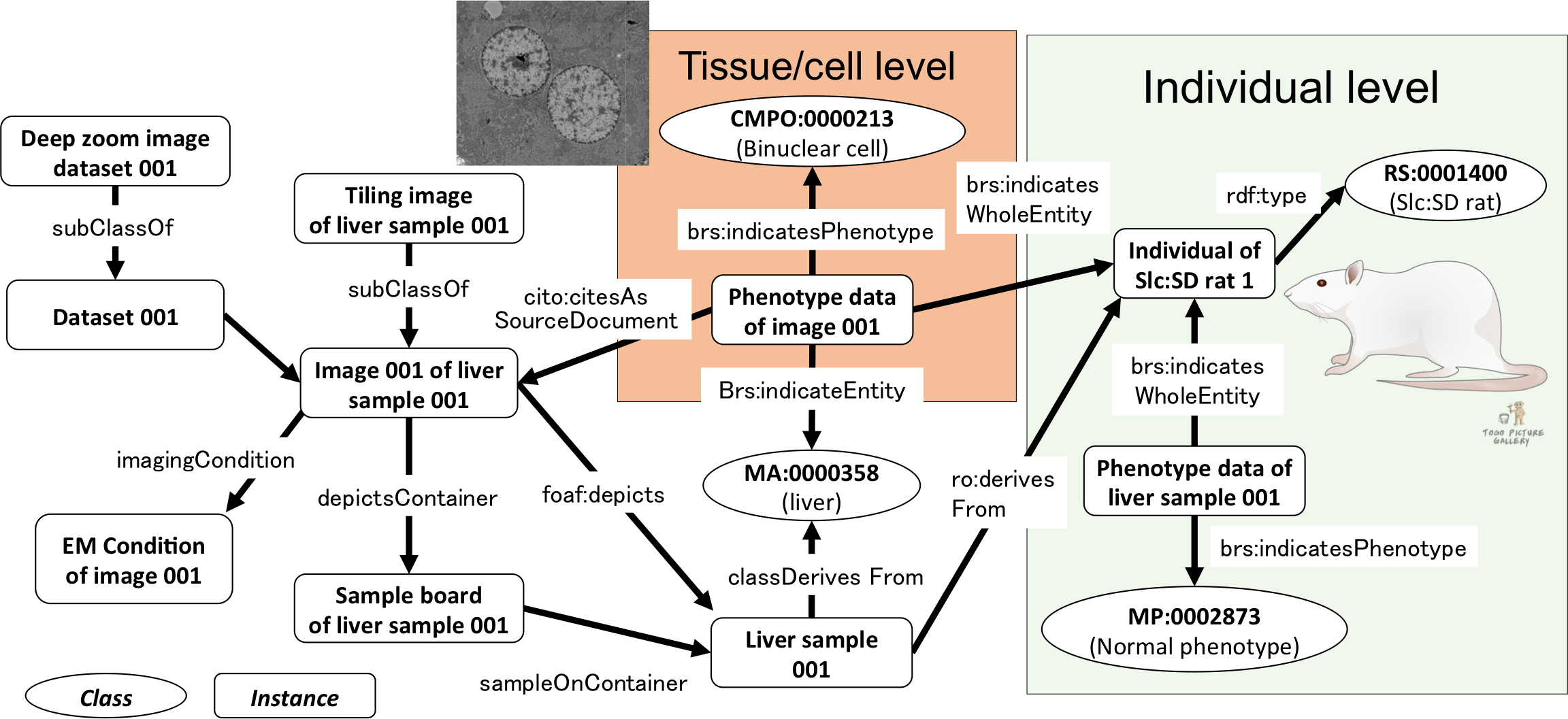}
\caption{Example graph structure showing that an abnormal binuclear phenotype of a liver cell is detected from the imaging analysis of a Slc:SD rat with a normal phenotype. The rounded rectangles are instances of classes, while ovals are classes. An EM image (Image 001 of liver sample 001) shows a liver sample (Liver sample 001) derived from a Slc:SD rat (Individual of Slc:SD rat 1). Referencing Image 001 of liver sample 001, phenotype data (Phenotype data of image 001) were produced. The graph structure can link bioresource information in RIKEN to an external database from RS:0001400.}
\label{fig:1}
\end{figure}

Further, we integrated such concepts which are having 15 super classes described using \texttt{rdfs:subClassOf} and 27 classes which defines choices using \texttt{rdf:type}. In this way, reusing data has progressed by handling such super classes and choice classes in a large scale, which was not organised by the XML-based OME data schema. Resultantly, imaging metadata for the rat liver tissue, such as imaging method, SEM parameters, experimental conditions, biological sample (a rat) and phenotype data, obtained from observations of the biosample and images were successfully described, as shown in Figure \ref{fig:1}.

\begin{figure}[htbp]
\centering
\vspace*{-2mm}
\includegraphics[width=10cm]{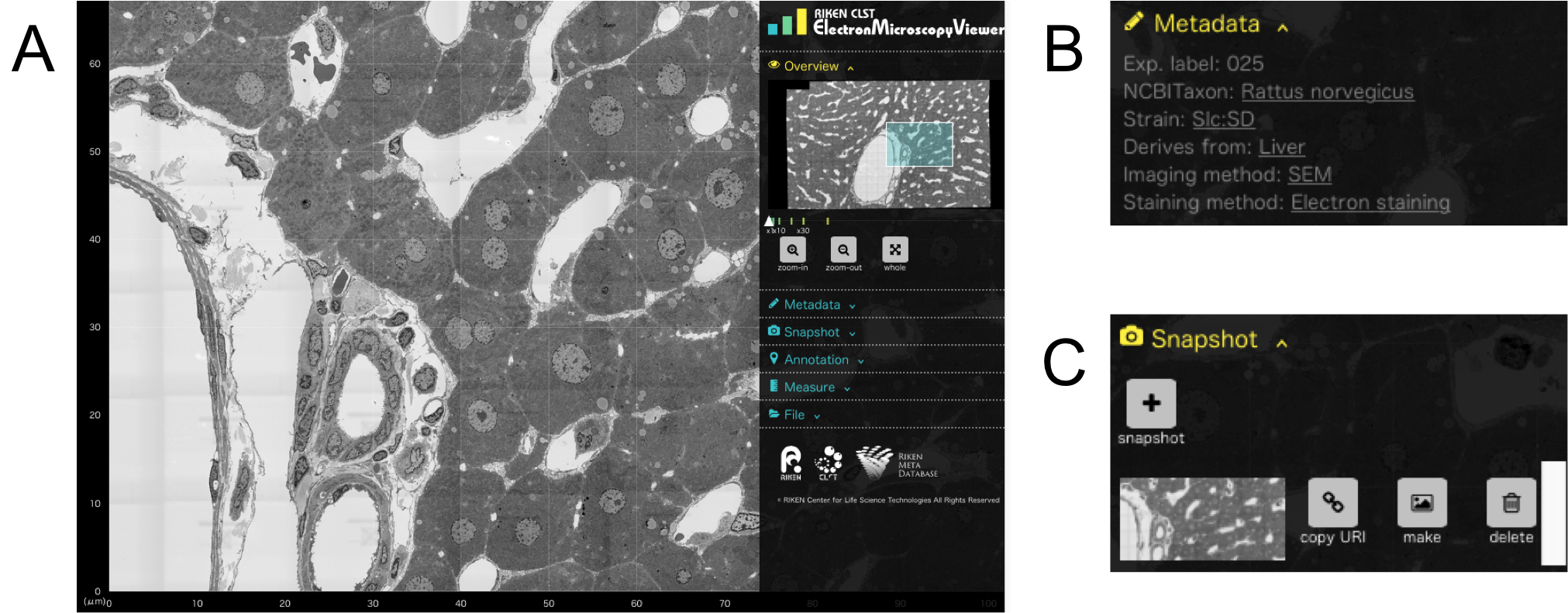}  
\caption{Interactive visualisation of large tiling image in the RIKEN CLST Electron Microscopy Viewer. (A) DZI of the tiling image visualised using the viewer and images around the interlobular vein of the liver. (B) The viewer contains metadata such as NCBI taxon, animal strain and derives from. (C) The viewer includes a snapshot function to capture a certain part of the image. These viewer data are published on the RIKEN web site (\url{http://clst.multimodal.riken.jp/CLST_ViewerData/EMV_025_161114SEM_RatLiverNormal/})}
\label{fig:2}
\end{figure}

\begin{figure}[htbp]
\centering
\vspace*{-2mm}
\includegraphics[width=11cm]{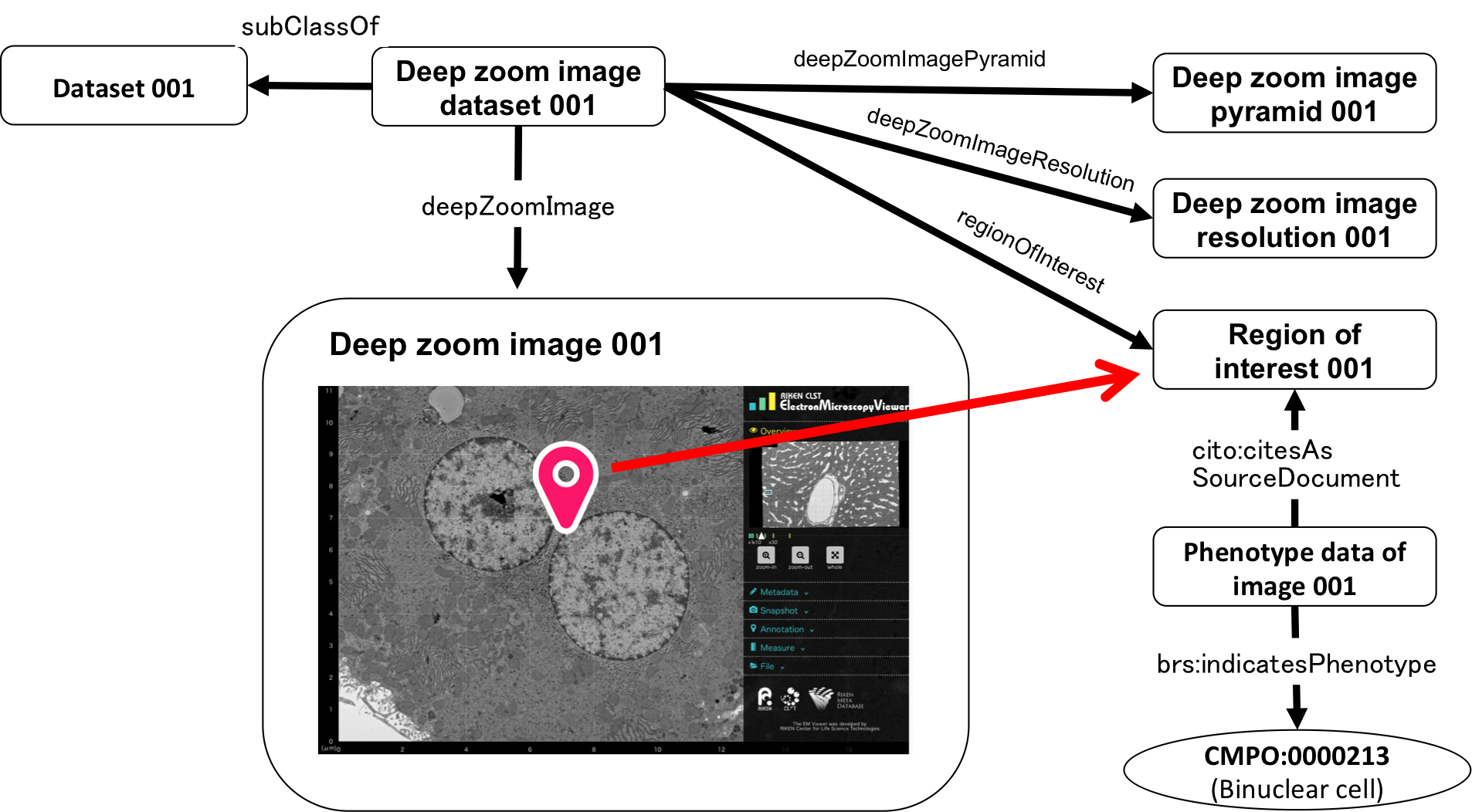}  
\caption{Metadata of a DZI based on the microscopy ontology to control the image viewer. The deep zooming function of the image viewer accesses the DZI dataset, which links the class of deep zoom image and other metadata components. The DZI dataset can have an ROI that can be linked to image phenotype annotations.}
\label{fig:3}
\end{figure}

On the other hand, we designed the semantics of the DZI dataset such that they can be handled properly by the image viewer application. We defined a DZI dataset to link to a DZI. By referring to the DZI URL, this viewer provides a graphical user interface to visualise the tiling image combined with hundreds of microstructural biotissue images (Figure \ref{fig:2}A). The viewer also functions successfully for the metadata card view for the tiling image (Figure \ref{fig:2}B) and generates snapshots and a direct URI for a part of an image (Figure \ref{fig:2}C). Furthermore, using the viewer’s annotation function, users can define a region of interest (ROI) and insert phenotype annotation of the cells or cellular components (i.e. binuclear cell phenotype (CMPO:0000213) of liver cells) (Figure \ref{fig:3}). Users can search the ROIs for a specific phenotype in the search mode by designating a phenotype term via a text search. Thus, the electron microscopy viewer is a powerful and a useful tool to accumulate phenotype information in morphome data. Also, the CMPO provided the description of the general phenotypic observations in various kinds of biological samples and utilised for phenotype annotation in several image repositories including the IDR, allowing for future cross-species integration of phenotypic data \cite{Jupp2016}.


\subsection{Imaging Metadatabase and Comprehensive Morphome Analysis on the Semantic Web}
Practically, a microscopy imaging database is required to provide various imaging metadata corresponding to various experimental results and conditions such as the disease phenotype of biotissues. Such metadata are a great benefit when users search images as linked data for research and education purposes. In this study, we conducted a series of histological experiments from the sample preparation of biotissue (rat liver) to imaging experiments by SEM. In addition, we described RDF-based metadata of these experimental processes using the microscopy ontology. We also verified that images, experimental conditions and phenotype data are linked and can be implemented as a practical metadatabase. Using the microscopy ontology, we have described metadata for approximately 300 images of rat liver tissue and have published the RIKEN Microstructural Imaging Metadatabase. As a result, we have successfully managed a vast number of images and their metadata, including the interpretation of morphological phenotypes occurring in sub-cellular components and biosamples captured in the images. Also, the accumulation of machine-readable knowledge in the morphome, e.g. metadata that include morphological phenotype data in diseases, can be utilised as a higher educational system for specialised experience as well as an advanced research in life sciences. Also by introducing such a semantic foundation to common medical imaging data such as CT, MRI and PET, which are done in the clinical field, the management and sharing of these imaging data will be considered to be beneficial.

\section{Conclusions}
We have developed a semantic web-based imaging database, i.e. the RIKEN Microstructural Imaging Metadatabase. As far as we know, a combination application of the ontology-based imaging metadatabase and its viewer is the first attempt in the world, and it has also made data integration with bioresources and other data in the life science possible. In this paper, we have shown an imaging dataset in the metadatabase that consists of approximately 300 images of rat liver tissue. To describe the imaging-related information, such as the experimental conditions of the EM and biosamples, we extended the original OME vocabularies for optical microscopy and used them to describe our EM imaging metadata. Moreover, we have developed an interactive EM image viewer that can visualise a tiling image in the DZI format by referring to the image metadata. We believe that these results will lead to the evolution of morphomics and emerging bioimaging research fields. Additionally, these results will be particularly useful for practical biomedical research and education.

\subsubsection*{Acknowledgements}
This paper is dedicated to RIKEN's centennial. This work was supported by the Management Expenses Grant for RIKEN CLST-JEOL Collaboration Center, the RIKEN Ageing project, and the RIKEN engineering network project.

%
%

%
\end{document}